\RequirePackage[l2tabu, orthodox]{nag} %

\documentclass[journal,twoside]{IEEEtran}

\usepackage{cite}

\usepackage{afterpage} %

\usepackage{amsmath}
\usepackage{amssymb}

\usepackage{graphicx}%
\graphicspath{{img/}} %
\usepackage{dcolumn}%
\usepackage{bm}%
\usepackage{siunitx} %

\usepackage{xcolor} %

\usepackage{hyperref} %
\usepackage{cleveref} %
\usepackage{url}

\usepackage[OT1]{fontenc}
\usepackage[strict=true]{csquotes} %
\usepackage[strings]{underscore} %

\hypersetup{
  colorlinks=true,
  linkcolor=cyan, citecolor=cyan, urlcolor=cyan,
  pdftitle={Drivers learn city-scale dynamic equilibrium},
  pdfauthor={Ruda Zhang and Roger Ghanem},
  pdfprintscaling=None,
}

\begin{document}
\bstctlcite{IEEEexample:BSTcontrol}

\title{Drivers Learn City-scale Dynamic Equilibrium}

\author{Ruda~Zhang and Roger~Ghanem%
  \thanks{Manuscript received...
    Research is funded by National Science Foundation (NSF) Grant No. 14-524
    Resilient Interdependent Infrastructure Processes and Systems and, in part,
    by NSF Grant DMS-1638521.
    \textit{(Corresponding author: Ruda Zhang.)}}%
  \thanks{R. Zhang and R. Ghanem are with the Department of Civil and Environmental Engineering,
    University of Southern California, Los Angeles, CA 90089 USA
    (e-mail: rudazhan@usc.edu; ghanem@usc.edu).
    R. Zhang is now with Duke University, Durham, NC 27710, USA.}}

\markboth{IEEE Transactions on Intelligent Transportation Systems,~VOL.~X, NO.~X, MONTH~202X}%
{Zhang and Ghanem: Drivers learn city-scale dynamic equilibrium}

\maketitle

\begin{abstract}
  Understanding driver behavior in on-demand mobility services
  is crucial for designing efficient and sustainable transport models.
  Drivers' delivery strategy is well understood,
  but their search strategy and learning process still lack an empirically validated model.  
  Here we provide a game-theoretic model of driver search strategy and learning dynamics, 
  interpret the collective outcome in a thermodynamic framework,
  and verify its various implications empirically. 
  We capture driver search strategies in a multi-market oligopoly model,
  which has a unique Nash equilibrium and is globally asymptotically stable.
  The equilibrium can therefore be obtained via heuristic learning rules
  where drivers pursue the incentive gradient or simply imitate others.
  To help understand city-scale phenomena,
  we offer a macroscopic view with the laws of thermodynamics.
  With 870 million trips of over 50k drivers in New York City,
  we show that the equilibrium well explains the spatiotemporal patterns of driver search behavior,
  and estimate an empirical constitutive relation.
  We find that new drivers learn the equilibrium within a year,
  and those who stay longer learn better.
  The collective response to new competition is also as predicted.
  Among empirical studies of driver strategy in on-demand services,
  our work examines the longest period, the most trips,
  and is the largest for taxi industry.
\end{abstract}

\begin{IEEEkeywords}
  on-demand service, driver strategy, search, learning, game theory, Nash equilibrium
\end{IEEEkeywords}

\section{Introduction}
\label{sec:introduction}

On-demand service is an increasingly important part of urban transportation,
and has significant impacts on the economy, congestion, and the environment.
Current on-demand mobility services include taxis and ridesourcing apps.
Both the street-hailing and the e-hailing models are competitive and complementary,
and will continue to co-exist \cite{NieY2017}.
Understanding the operation of on-demand transport can guide regulations on the current market,
and help us design more efficient and more sustainable service models in the future
\cite{Egan2016, Alonso-Mora2017, Vazifeh2018, Ramezani2018}. %

With global positioning system (GPS) trajectory data,
researchers have been able to analyze the spatial-temporal demand-supply patterns of
on-demand services over the last decade, see \cite{WangSF2019} for a review.
Of particular interest are driver behaviors.
Drivers for on-demand services work as independent contractors,
who move around a city to maximize income.
We want to understand what determines driver behavior,
and how drivers learn to adjust their behavior.

Current studies of driver behavior in on-demand services
have formulated driver strategy in various ways.
Overall, the main strategic factors include
driving speed (in both delivery and search), search location, and driver experience.
Before GPS data were available,
\cite{Schaller2007} studied taxi markets in 43 cities and counties in North America,
and found that drivers cluster in locations with high trip density,
regardless of entry policy and license management.
Perhaps the first study of driver strategy using GPS data is \cite{LiuL2010a},
which finds that taxis with high daily revenue are strategic
in choosing areas to serve at different time of a day to avoid traffic and competition.
\cite{LiB2011} studied driver search strategies at different time and location,
and concluded that searching is always better than waiting.
A follow-up work \cite{ZhangDQ2015} extended the driver strategy vector to
include delivery speed and service-region preference (proportions of search trips across areas),
and found effective strategies in each factor.
\cite{QinGY2017} fit an econometric model of driver income
with five strategic factors that may explain income difference,
and found that the significant factors are (in descending order of contribution)
delivery speed, search distance, supply-demand ratio, and trip fare,
while search intensity is not statistically significant.
\cite{TangLL2017} found that taxis with high single-trip efficiency,
defined as the average income rate between two consecutive pickups,
usually avoid traffic and seek locations with high passenger demand.
Among studies of this kind, \cite{Cook2020} is perhaps the largest in scale to date.
Using operational data on 1.87 million Uber drivers in the US over two years,
they find that the income difference between male and female drivers
can be completely explained by three factors (relative contribution in parenthesis):
delivery speed (1/2), experience on the platform (1/3),
and preferences over where and when to work (1/6).
Specifically, drivers earn more if they drive faster on trip, have more experience,
and drive in locations with lower wait times and higher surge multipliers.

Other driver decisions can affect income as well,
such as driving safety \cite{Jackson2011},
when to start and stop working \cite{Farber2015,Frechette2019},
detours and overcharges \cite{Balafoutas2013, Balafoutas2015},
and passenger denial \cite{ZhangSH2016}.
But based on the results of \cite{Cook2020},
we can focus on delivery speed, search strategy and experience.
We note that per \cite{Cook2020}, delivery speed is only slightly increasing in experience,
but experience contributes to search strategy and other minor strategic factors.
Therefore, search strategy and experience together are as important as delivery speed
for driver income.

Fast delivery speed is an important factor that increases driver income,
which has been unambiguously supported by the research reviewed so far.
On the topic of delivery route choice,
although there are seemingly conflicting results,
the consensus is that high-income drivers take faster routes.
For example, \cite{SunJ2014} and \cite{Yildirimoglu2018}
found that most drivers do not travel along the fastest (or the shortest) paths.
However, \cite{LiuL2010a} and \cite{DuanZY2014} found that
high-income drivers can find a faster and more reliable route than low-income drivers.
Furthermore, \cite{LiL2018} found that taxi drivers usually choose one of the few fastest routes.
An alternative theory for route choice is that drivers anchor towards urban features on the path,
and \cite{Manley2015} gave empirical evidence for it.
However, the data is from a prearranged service,
which has a different incentive structure than the on-demand services we mentioned earlier.

Search strategy is more complex and, in a sense, more interesting.
Current models of driver search strategy use decision theory or game theory.
Both \cite{Wong2015} and \cite{TangJJ2016} used two-stage discrete-choice models,
such that a driver first chooses the next pickup location after finishing a trip,
and then chooses a route to get there.
In \cite{Wong2015}, an urban area is divided into zones
and each zone is divided into a network of cells.
The first stage uses a multinomial logit model for the choice of a target zone,
and the second stage uses a sequence logit model for the sequential choice of cells \cite{Wong2014}.
In \cite{TangJJ2016},
the first stage uses a Huff model to describe the probability distribution of pickup locations,
which combines location attractiveness and travel cost;
the second stage uses a path size logit model to find a path on the road network
to the target location, considering path size, path distance, travel time, and intersection delay.
For the decision on whether to return to an airport which is distant but has high trip values,
\cite{Yazici2016} used a logit model and found that the most essential factors
include the drop-off location and whether the driver has a ``short return ticket''.
To account for uncertainties in travel time, waiting time, and passenger travel distance
at high-value pickup locations such as airports and train stations,
\cite{ZhengZ2018} proposed a model where drivers
have their subjective beliefs on such uncertain values,
make decisions to maximize expected reward over a time horizon,
and learn from past experiences by Bayesian updating of beliefs.
Their simulation results show that this Bayesian learning process
is effective in increasing the subjective reward.

Game theoretic models explicitly consider the search strategies of other drivers,
and predict driver strategies as a non-cooperative equilibrium.
For a static game with driver--customer bilateral search and elastic demand,
\cite{YangH2010} proved the existence of Nash equilibrium (NE)
and provided a solution algorithm via best-response dynamics.
In a follow-up work, \cite{YangH2011} studied the comparative statics of taxi equilibrium:
with variable taxi fleet size and trip fare,
how does the equilibrium outcome change in terms of taxi utilization rate and customer wait time,
and which combinations maximize taxi profit and social welfare
(conditions known as monopoly optimum and social optimum, respectively).
To account for e-hailing service,
\cite{HeF2015} extended the game model of \cite{YangH2010} and provided similar results.
Modeling the same problem with a multi-leader follower game,
\cite{QianXW2017} proved the existence of generalized NE, and provided a solution algorithm
that iteratively solves the augmented variational inequalities of the leader and the follower.
All four studies used numerical simulations instead of empirical data.
\cite{Buchholz2021} used a dynamic spatial oligopoly model for a taxi market,
where drivers choose search locations over a finite time horizon,
and are assumed to know search location profitability and competitor locations.
He proved that there exists a unique, symmetric equilibrium
for the spatial-temporal distribution of vacant taxis and passenger demand.

Driver experience and learning is a less studied topic.
Besides \cite{Cook2020} and \cite{ZhengZ2018},
\cite{Haggag2017} showed that neighborhood-specific local experience has a significant impact
on drivers' search strategies following drop-offs;
and \cite{ZhangYJ2020} showed that high-income drivers benefit significantly
from their ability to learn from local and global demand information.

\begin{table}[tb]
  \centering
  \caption{GPS trajectory data of on-demand services\\
    used in studies of driver strategy: an incomprehensive list.}
  \begin{tabular}{crrrrr}
    \hline
    Reference            & Location  & Period          & Trips & Vehicles & Drivers \\
    \hline                                                     
    \cite{LiuL2010a}     & Shenzhen  & 2008            & -     & -        & 3k      \\
    \cite{ZhangYJ2020}   & Shenzhen  & 2009-09         & -     & -        & 11.2k   \\
    \cite{LiB2011}       & Hangzhou  & 2009-10 (15d)   & -     & 4,548    & -       \\
    \cite{ZhangDQ2015}   & Hangzhou  & 2009-04:2010-03 & -     & 7,600    & -       \\
    \cite{ZhengZ2018}    & Guangzhou & 2009-05-11      & 24k   & 758      & -       \\
    \cite{Wong2015}      & Hong Kong & 2009-08-16:23   & -     & 460      & -       \\
    \cite{QinGY2017}     & Shanghai  & 2011-05 (20d)   & -     & 8,000    & -       \\
    \cite{TangLL2017}    & Wuhan     & 2013-08-04:10   & -     & 2,000    & -       \\
    \cite{TangJJ2016}    & Beijing   & 2014-09:2015-02 & -     & -        & 36k     \\
    \cite{Yazici2016}    & NYC       & 2009-03:05      & 44.5m & -        & -       \\
    \cite{Haggag2017}    & NYC       & 2009            & 171m  & -        & 6.3k    \\
    \cite{Frechette2019} & NYC       & 2011-2012       & -     & -        & 37.4k   \\
    \cite{Buchholz2021}  & NYC       & 2012-08:09      & 27m   & -        & -       \\
    \cite{Cook2020}      & US        & 2015-01:2017-03 & 740m  & -        & 1.87m   \\
    This paper           & NYC       & 2009-2013       & 870m  & 13,237   & 50.3k   \\
    \hline
  \end{tabular}
  \label{tab:data}
\end{table}

In this paper, we use a game-theoretic model
to predict drivers' spatio-temporal search strategies,
examine the individual and group learning dynamics as drivers gain experience,
and validate these model predictions with large-scale GPS trajectory data.
Following our initial work \cite{ZhangRD2018phd,ZhangRD2020dsp},
we regard the regularity in urban transportation
as the equilibrium outcome of individual decision-making,
in response to transportation demand and services.
For on-demand services,
given exogenous traffic speeds and passenger demand,
the income maximization of drivers leads to an economic equilibrium.
We formalize drivers' decision-making as a non-cooperative game (\cref{eq:game-simplified,eq:game}),
which has a unique Nash equilibrium that is stable under simple learning dynamics
such as adaptive learning and social learning \cite{ZhangRD2020game}.
Besides this microscopic individual-level model,
we also provide an interpretation of the equilibrium in a macroscopic thermodynamic framework,
and describe the laws of thermodynamics (\cref{eq:zeroth-law,eq:first-law,eq:second-law}),
constitutive relation (\cref{eq:constitutive}),
and fundamental thermodynamic relation (\cref{eq:fundamental}).
With five years of New York City (NYC) taxi trip records \cite{ZhangRD2019data},
we validate the equilibrium in space (\cref{fig:validation}a-b)
and over time (\cref{fig:validation}c),
and estimate an empirical constitutive relation (\cref{fig:validation}d).
We also examine the learning process of individual drivers (\cref{fig:learning}),
and apply the thermodynamic framework to predict
drivers' adjustment as a group to a new system (\cref{fig:adjustment}).
We discuss the economic efficiency of taxi transportation,
and show evidence (\cref{fig:markov}) and equivalence (\cref{eq:equivalence})
of a Markovian formulation of driver search strategy.%

The main contributions of our paper are summarized as follows.
(1) This is the first large-scale study of driver search strategy and learning,
which complements \cite{Cook2020}.
(2) We provide a game theoretic model that is empirically validated;
such models of driver behavior are not found in the literature except \cite{Buchholz2021}.
(3) We give evidence of Markovian search strategy,
which is often hypothesized but not validated in the literature.
(4) Among empirical studies of driver strategy in on-demand services,
our work examines the longest period, the most trips,
and is the largest for the taxi industry (see \Cref{tab:data}).%

\section{Theoretical Results}

\subsection{Game model}
\label{sub:game}

We assume that each driver chooses their driving strategy to maximize income,
and we define the strategy of a driver in service to be
how they allocate their service time across the city (\cref{fig:overview}a-b).
Let $E$ be the set of street segments in the road network
and $N$ be the set of drivers in service.
Let $s_{ix}$ be the proportion of service time driver $i$ allocates on street segment $x$,
then the driver's strategy is $\mathbf{s}_i = (s_{ix})_{x \in E}$,
which sums to 1 (\cref{fig:overview}c).
We can formalize a driver's decision as an optimization problem
(see \cref{sub:decision} for details):
\begin{equation}
  \label{eq:game-simplified}
  \begin{aligned}
    & \text{maximize}    && \pi_i(\mathbf{s}_i; \mathbf{s}_{-i}, \mathcal{E}) \\
    & \text{subject to}  && \mathbf{s}_i \ge 0 \\
    &                    && \mathbf{s}_i \cdot \mathbf{1} = 1
  \end{aligned}
\end{equation}
Here, with hour as the unit of time, $\pi_i$ is the expected hourly revenue of the driver,
and $\mathbf{s}_{-i} = \sum_{j \ne i} \mathbf{s}_j$ is the aggregate strategy of other drivers.
Environment condition $\mathcal{E}$ includes traffic speed and passenger demand,
and is regarded as constant during short time windows of analysis.

Competition among drivers could lead to specific choices of strategies, called equilibrium.
If we see every street segment as a distinct market
and every driver in service as a multi-market firm,
we can abstract \cref{eq:game-simplified} as a game of
multi-market competition among firms of equal capacity \cite{ZhangRD2020game}.
This game has a unique Nash equilibrium, %
where all drivers use the same strategy
and marginal driver revenue are uniform across all searched segments.
Moreover, the equilibrium is globally asymptotically stable
under adaptive learning \cite{Arrow1960} %
and/or imitative learning \cite{Pentland2014,Roth1995, Fudenberg2009}.
Denote this equilibrium as $S^* = (\mathbf{s}_i^*)_{i \in N}$,
where $N$ is the set of drivers in service.
Because all drivers use the same strategy, let $\mathbf{s} = \sum_{i \in N} \mathbf{s}_i$,
we can write $S^* = n^{-1} \mathbf{s}^* \mathbf{1}_n^{\text{T}}$,
where $n = |N|$ is the number of drivers in service.
The equilibrium can be determined such that
$\mathbf{s}^*$ is the unique point that maximizes a potential function:
\begin{equation}
  \Phi(\mathbf{s}) = \sum_{x \in E} \int_0^{s_x} \phi_x(t)~\mathrm{d}t
\end{equation}
where $\phi_x(s_x^*) = (\partial \pi_i / \partial s_{ix}) (S^*)$
is marginal driver revenue on a segment at equilibrium,
and $s_x = \sum_{i \in N} s_{ix}$, see~\cref{fig:overview}d.

\subsection{Thermodynamic interpretation}
\label{sub:thermodynamics}

We can interpret the Nash equilibrium %
as a thermodynamic equilibrium.
This establishes a macroscopic equilibrium where aggregate behavior is perceived as
a transport phenomenon built up from individual choices.
This macroscopic view ignores the decision-making and competition of drivers,
but helps understand the outcome of a social system from the perspective of a physical system.

We regard drivers as interchangeable particles with identical behavior at equilibrium.
Regard total service time $s$, which equals the number of drivers in service,
as total energy of the taxi transportation system.
Regard potential function $\Phi$ of the game as entropy of the system. %
And regard the reciprocal of equilibrium marginal driver revenue, $\psi = 1 / \phi$, as temperature.
Then $s$, $\psi$, and $\Phi$ are all state variables of the system at equilibrium
given environment condition $\mathcal{E}$.

Being a state variable and intensive property,
temperature $\psi$ is the driving force of
the transport of service time $s$ across the street segments.
As we mentioned earlier, the learning process of the game %
always increases the potential function $\Phi(\mathbf{s})$, which is maximized at equilibrium.
When two systems at equilibrium are put into contact
with an interface permeable to the transfer of service time,
$s$ will flow from the system with higher $\psi$ to the one with lower $\psi$.
At equilibrium, $\psi$ is uniform across all searched segments.
In summary, we can make the following statements of thermodynamics.
Zeroth law: two taxi systems in contact have the same equilibrium marginal driver revenue.
\begin{equation}
  \label{eq:zeroth-law}
  \psi_1 = \psi_2
\end{equation}
First law: taxi transportation is the transfer process of total service time $s$,
which is a conserved quantity.
\begin{equation}
  \label{eq:first-law}
  \mathrm{d} s = \sum_{x \in E}{\mathrm{d} s_x}
\end{equation}
Second law: under fixed demand and traffic state,
a closed taxi system maximizes its potential function. %
\begin{equation}
  \label{eq:second-law}
  \mathrm{d} \Phi \ge \frac{\delta s}{\psi}
\end{equation}

Zeroth law defines equivalent classes of equilibrium,
which are strictly totally ordered by state variable $\psi$.
The manifold of equilibrium is thus one-dimensional, parameterized by $\psi$,
and any other state variable must depend on it.
This means that state space $(\Phi, s, \psi) \mid \mathcal{E}$ has only one degree of freedom,
and this dependency is the constitutive relation of the system
given environment condition $\mathcal{E}$,
with explicit form
\begin{equation}
  \label{eq:constitutive}
  (\Phi, \psi)(s) \mid \mathcal{E}
\end{equation}

Rearranging the exact differential of $s(\Phi) \mid \mathcal{E}$
gives the fundamental thermodynamic relation of the equilibrium:
\begin{equation}
  \label{eq:fundamental}
  \mathrm{d}s = \psi \mathrm{d}\Phi
\end{equation}

We test various implications of this theory of thermodynamics in our empirical results.

\section{Empirical Results}

\afterpage{%
\begin{figure*}[tb]
\centering
\includegraphics[width=\linewidth]{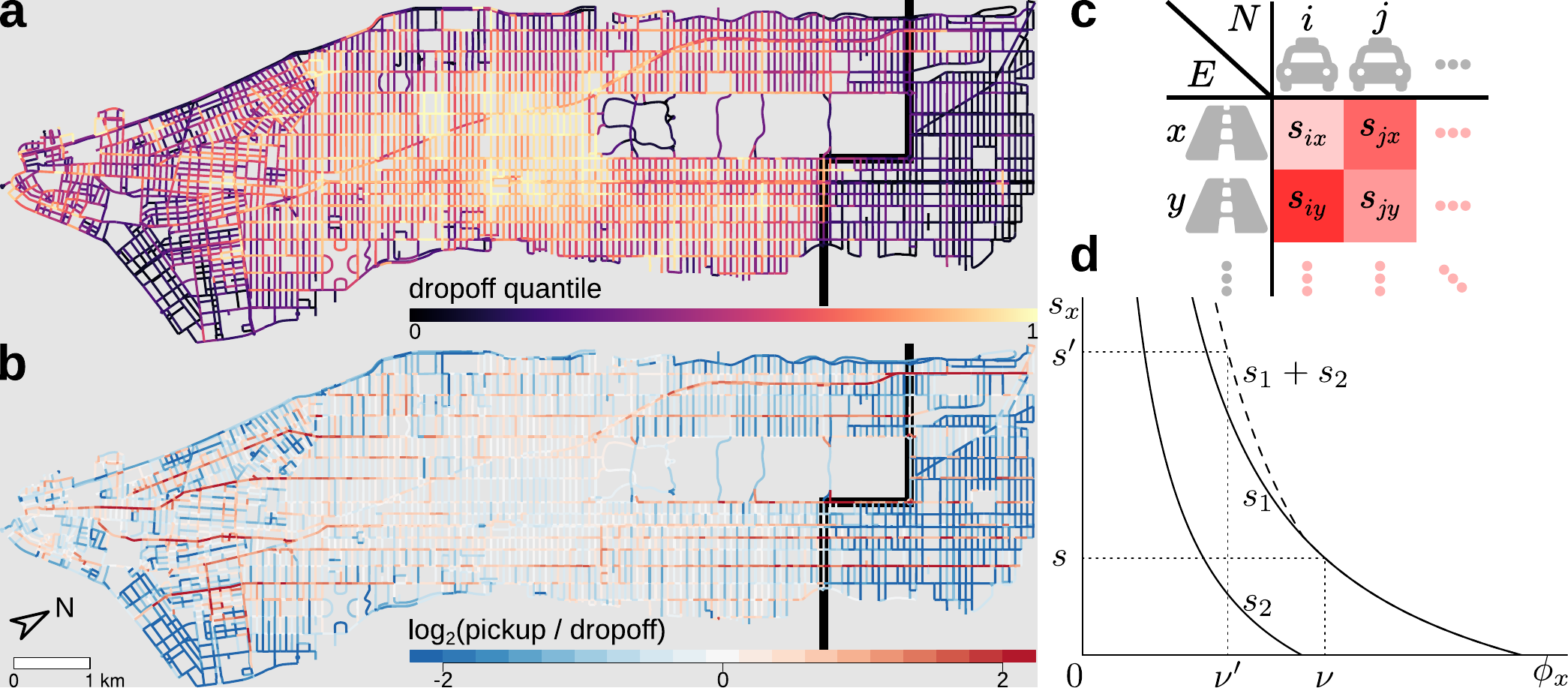}
\caption{
  Driver strategy.
  \textbf{a-b}, Manhattan street network used in this paper,
  showing characteristics of taxi activity:
  \textbf{a}, quantile of drop-off per segment length;
  \textbf{b}, log2 pickup-dropoff ratio.
  Black bold line marks the north border of core Manhattan.
  \textbf{c}, drivers allocate their service time across the segments, which can differ.
  \textbf{d}, determination of equilibrium.
  With $s$ drivers, marginal driver revenue $\nu$ on a segment at equilibrium
  is determined by $\sum_{x \in E} s_x(\nu) = s$.
  Equilibrium allocation on each segment can then be determined by $s_x^* = s_x(\nu)$.
}
\label{fig:overview}
\end{figure*}

\begin{figure*}[t!]
\centering
\includegraphics[width=\linewidth]{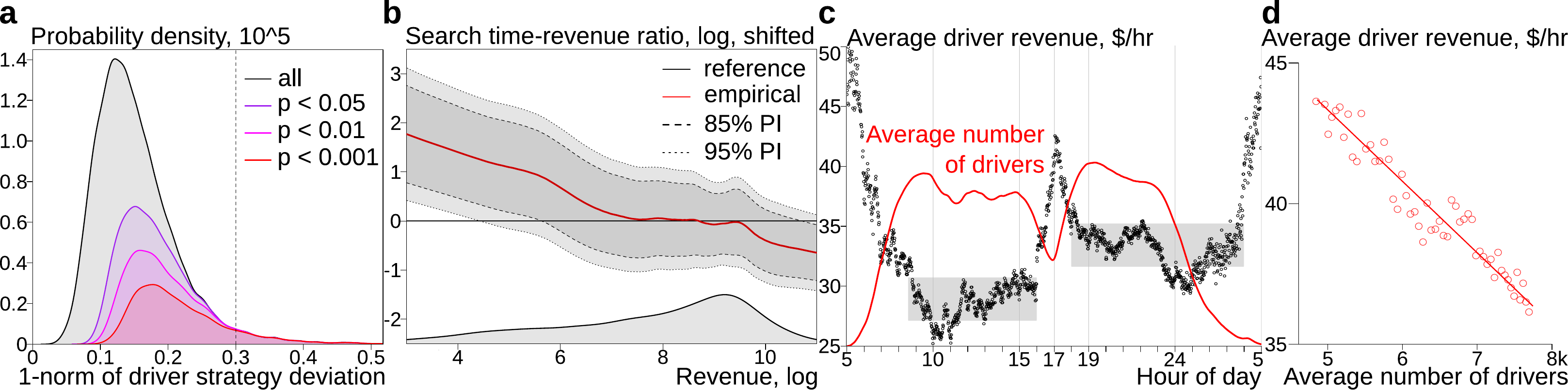}
\caption{Verification of equilibrium.
  Using trip records in Spring 2011 and Spring 2012.
  \textbf{a}, probability distribution of 1-norm of driver deviation from average strategy,
  in Tue-Thu PM peaks, 6pm--10pm, grouped by $p$-values.
  3.66\% of drivers have statistically significant ($p > 0.05$) large deviations
  ($\|\mathbf{x}'\|_1 > 0.3$).
  \textbf{b}, log of search time--revenue ratio on street segments, Mon-Fri 6pm--7pm,
  shifted to a reference value. Local regression (red); prediction intervals (shade);
  distribution of log revenue on segments (margin).
  \textbf{c-d}, average number of drivers and driver revenue, Wed 5am--Thu 5am,
  each dot represents one minute: %
  \textbf{c}, time series, rectangles mark AM shift (8:30am--4pm) and PM shift (6pm--4am);
  \textbf{d}, trajectory during 5pm, red line shows a linear regression.
}
\label{fig:validation}
\end{figure*}

\begin{figure*}[t!]
  \centering
  \includegraphics[width=0.96\linewidth]{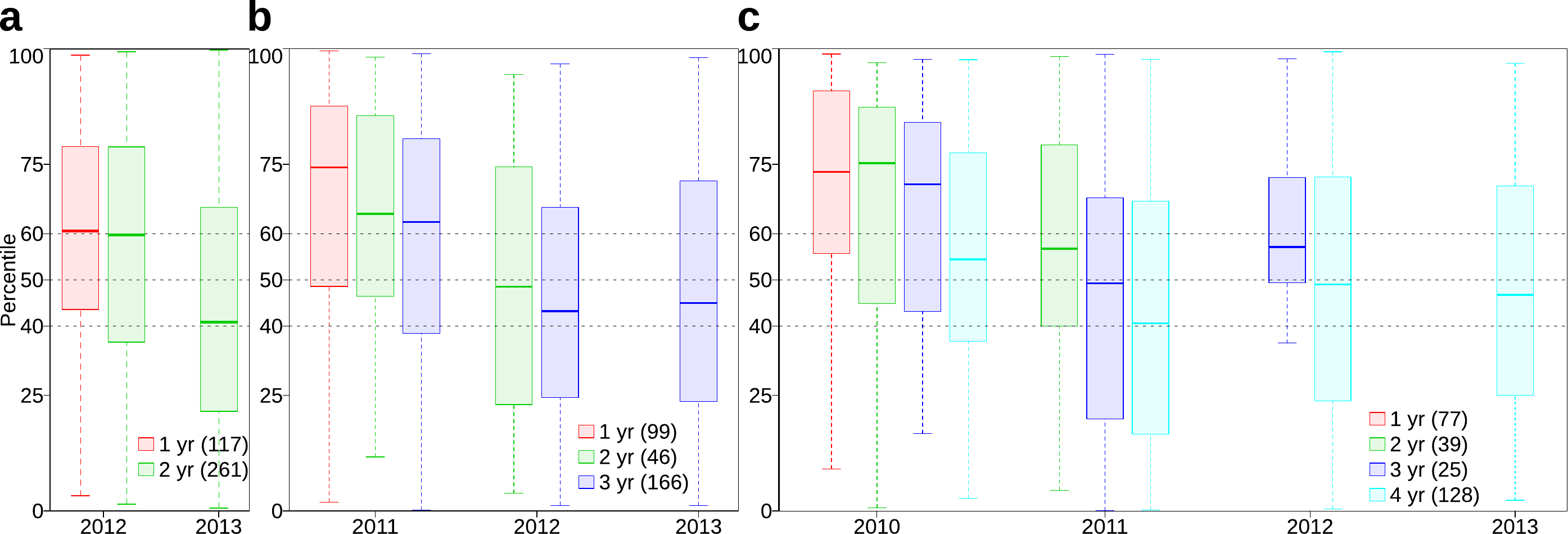} %
  \caption{Individual learning of equilibrium.
    Percentile of strategy deviation of new drivers joined in
    \textbf{a} 2012, \textbf{b} 2011, and \textbf{c} 2010,
    grouped by years of consecutive driving up to 2013,
    size of each group shown in parentheses.
    Using trip records in springs, Tue-Thu PM peaks, 6pm--10pm.
  }
  \label{fig:learning}
\end{figure*}

\clearpage
}

\subsection{Verification of spatial equilibrium}
\label{sub:validation}

To verify that drivers actually follow the theoretical equilibrium, we proceed in two parts:
(1) all drivers shall use the same strategy;
and (2)
marginal driver revenue must be uniform across all searched segments.

Although driver search strategy is not directly observed,
it is proportional to driver pickup probability on each segment (\cref{eq:prop-pickup-search}).
If all drivers use the same strategy,
each driver's pickup probability distribution across segments
shall be the same as that of the overall distribution.
To test this implication,
we partition the segments into 10 equi-probable groups, %
then a random pickup can be modeled as a categorical random variable,
with equal probability $p = 0.1$ for each group.
As a sum of such categorical trials,
each driver's pickup count can then be modeled as a multinomial random variable
with $p = 0.1$ for each group.
We test drivers' pickup counts in these groups by a corrected log likelihood ratio
of multinomial distributions \cite{Smith1981}.
For each driver, the pickup counts are normalized into a probability vector
$\mathbf{x} = (x_1, \dots, x_{10})$, which is then summarized by the 1-norm $\|\mathbf{x}'\|_1$
where $\mathbf{x}' = \mathbf{x} - \mathbf{1}/10$.
Note that $\|\mathbf{x}'\|_1 = 0.2$ if the pickups in any one group
is arbitrarily allocated to the other groups;
and we consider a strategy to be a large deviation if $\|\mathbf{x}'\|_1$ exceeds 0.3.
\Cref{fig:validation}a shows the distribution of $\|\mathbf{x}'\|_1$.
Only 3.66\% of drivers have statistically significant large deviations.
Regardless of the arbitrary threshold for large deviation,
the result shows that most drivers use similar strategies.
Therefore we can think of drivers as particles with identical behavior.

Now we verify that all segments have the same marginal driver revenue,
or equivalently, the same temperature, as stated in the zeroth law \cref{eq:zeroth-law}.
We note that when $n \gg 1$, $\phi_x \approx \pi_x / s_x$,
where $\pi_x$ is the revenue originated on a segment
and $s_x$ is the total service time attributable to the segment.
Because at any moment the number of drivers in service in Manhattan is in the thousands,
this approximation is suitable.
So it suffices to show that $\pi_x$ is proportional to $s_x$,
which is the sum of search time $t_{sx}$ and trip time $t_{px}$ per unit time.
Because the majority of trips are metered,
which is calculated from trip distance and time in slow traffic,
driver revenue from each trip is highly correlated to trip duration regardless of driver strategy,
especially when traffic speed is hold stationary.
To avoid the influence of this fact,
consider trip time as a linear function of trip revenue,
then $\pi_x \propto s_x$ is equivalent to $\pi_x \propto t_{sx}$,
and we try to show the latter.
Because search routes are not recorded in the trip records,
we take trip records between 6pm and 7pm on weekdays, %
and estimate search routes between trips by shortest distance routing.
We consider this approach acceptable because during the selected hours,
traffic is roughly at a uniform congested speed
while average search time is the shortest (3 min, 80\% up to 6 min; see \cite{ZhangRD2018phd}),
so route deviation from the shortest path is unlikely.
\Cref{fig:validation}b shows $\log(\tilde{t}_{sx} / \pi_x)$ versus $\log(\pi_x)$,
where $\tilde{t}_{sx}$ is the estimated search time.
The majority of street segments have similar search-revenue ratios,
while segments with low revenue appear to be over-supplied
and those with very high revenue under-supplied.
We note that, for segments with low revenue,
marginal driver revenue might not be equilibrated
since they contribute little to driver revenue.
Our estimation assigns search time equally to each segment on route,
which may underestimate the actual search time near the pickup location,
and therefore underestimate search time on high revenue segments.
Moreover, shortest path routing provides a single route
for trips with the same origin and destination,
so the estimated search time may be concentrated on a few street segments,
which contributes to estimation error.
Therefore, accounting for these factors,
temperature $\psi$ is approximately uniform over space.

\subsection{Equilibrium dynamics}
\label{sub:dynamic}

As environment condition $\mathcal{E}$ varies over times of a day, the equilibrium will also vary.
If drivers are free to choose when to work
and are indifferent about working at different times of a day,
by zeroth law \cref{eq:zeroth-law}, driver supply $s$ will adjust so that %
temperature $\psi$ is stationary throughout a day.
Equivalently, $\phi$ stays the same throughout a day.
Note that marginal driver revenue on a segment and average driver revenue
are approximately the same at equilibrium:
because $\phi_x \approx \pi_x / s_x$, therefore $\phi \approx \sum_x \pi_x / \sum_x s_x = \pi / s$.
This means that, given the assumptions, average driver revenue is the same throughout a day.
To verify this, we examine the trajectory of average driver revenue
and number of drivers throughout a typical weekday, shown in \cref{fig:validation}c.
Average driver revenues during 8:30am--4pm and 6pm--4am
center around \$29/hour and \$33.5/hour respectively,
and are constant in the sense that its overall variation is about the same
as its short-term variation.
The difference between average driver revenue for these two periods can be explained by two factors.
First, the total number of taxis is limited and not all is available for the night shift,
so not all drivers who would like to work at night can get a taxi.
Second, the lease rate for day shifts is less than those of night shifts,
so the difference in average driver income between the two periods
is less than that of average driver revenue.
During 4pm--6pm most double-shifted taxis change drivers,
which means supply decisions during this period is not up to the drivers,
so the average driver revenue is not constant.
During 3am--6am very few drivers are at work,
and the high average driver revenue justifies the cost of working
when most people prefer to be sleeping.
During 6am--8:30am most day shift drivers start working,
and although the average driver revenue is not constant,
it stabilizes as more drivers become active.

In contrast to the equilibration of average driver revenue over time,
by constitutive relation \cref{eq:constitutive}, marginal driver revenue on a segment at equilibrium
is a decreasing function of the number of drivers given environment condition:
$\phi(s) \mid \mathcal{E}$.
This constitutive relation is hard to measure without controlled experiments,
but can be measured from observational data
if the number of drivers is forced to change much faster than the environment does,
such as during shift transition.
In \cref{fig:validation}d, the downward trend reflects $\phi(s)$ for 5pm--6pm,
when people leave work and taxis return for the night shift.
We see that, with a thousand more drivers in service, average driver revenue decreases by \$2.53/hour.

\subsection{Individual learning}
\label{sub:learning}

It is natural to ask if drivers learn to use the same strategy
that results in a spatially uniform marginal revenue.
We use drivers' first appearance in trip records to infer if they are new or experience drivers.
The rate of new drivers stabilizes around September 2009, with about 10.23 new drivers each day since.
For new drivers joined each spring from 2010 to 2012,
we compute the 1-norm of their strategy deviation, $\|\mathbf{x}'\|_1$,
and compare it with the group of experienced driver who worked through 2010-2013.
In particular, we group each year's new drivers
by their eventual consecutive years of driving up to 2013,
and track their percentile of $\|\mathbf{x}'\|_1$ against the experienced drivers.
\Cref{fig:learning} provides box plots for the groups.
Note that the experienced drivers, if plotted,
would always have the median and the first and third quartiles at 50, 25, and 75, respectively.
For all groups of new drivers who stayed for at least a year,
their strategy deviation decrease significantly in the second year,
with the median reducing between 10 to 20 percentile.
For new drivers who stayed through 2013 and for at least two years,
their strategy deviation stabilize in the later years
and are smaller or the same as the experienced drivers.
Moreover, new drivers who stay longer always have smaller strategy deviations than their cohorts.
We see that new drivers learn the equilibrium strategy within one year of driving.

\subsection{Group adjustment}
\label{sub:policy}

\begin{figure}[tb]
  \centering
  \includegraphics[width=\linewidth]{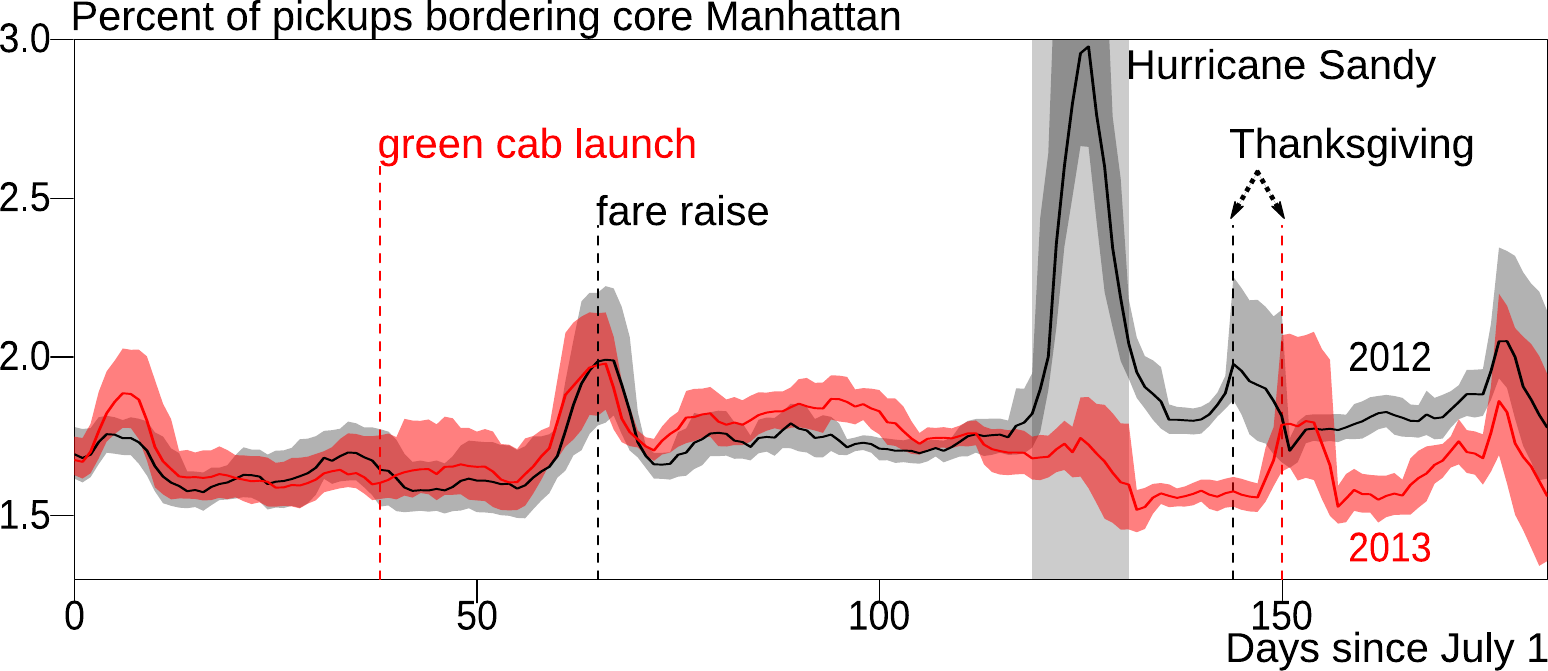}
  \caption{Group adjustment of equilibrium.
    Pickup probability in the region bordering core Manhattan,
    in the second halves of 2012 (black) and 2013 (red),
    7-day rolling value with 90\% bootstrap confidence band.
    Significant events and period marked by dashes and shade.
  }
  \label{fig:adjustment}
\end{figure}

Changes in taxi regulation affect the equilibrium,
which provide unique opportunities to test the implications of the theory.
On 2013-08-08, NYC TLC launched Street Hail Livery, also known as green cabs.
The new system is allowed to pick up street-hail passengers outside core Manhattan,
defined as south of West 110th Street and East 96th Street, see \cref{fig:overview}a-b.
This change gradually increased the supply of street-hail service outside core Manhattan,
and by second law \cref{eq:second-law} and constitutive relation \cref{eq:constitutive}
this should decrease the marginal driver revenue on segments therein.
By zeroth law \cref{eq:zeroth-law}, segments within core Manhattan
should also have marginal driver revenue decreased to the same level,
which implies more supply of yellow cabs in core Manhattan
where they have exclusive rights to service.
By first law \cref{eq:first-law},
the proportion of service time yellow cab drivers spent outside core Manhattan should decrease.
\Cref{fig:adjustment} compares the time-series of percentage of pickups
in the region bordering core Manhattan in 2012 and 2013.
This percentage slightly reduced after the 2012 fare raise,
greatly increased during Hurricane Sandy,
and moderately increased during Thanksgiving and Christmas.
Excluding irregularities due to Hurricane Sandy and the holidays,
the percentage is stable in the last two months of both years,
with a robust decline in 2013. %
This is consistent with the implication of our model,
and exemplifies the use of the thermodynamic interpretation of the equilibrium
in explaining macroscopic phenomena.

\section{Discussion}
\label{sec:Discussion}

\subsection{Economic inefficiency of the equilibrium}
\label{sub:inefficiency}

In multi-market oligopoly of equal capacity,
the Nash equilibrium is not economically efficient in general,
that is, the total income is not maximized \cite{ZhangRD2020game}.
For taxi industry, this means that the decentralized optimization of income
by self-interested drivers does not optimize total income of the industry.
Maximum total income may be achieved if drivers' search behavior is directed by a central planner.
Considering that ridesourcing apps match passengers with drivers,
this optimum may be attainable.
However, drivers can still have a spatio-temporal search strategy
and can cancel and accept trips strategically \cite{Cook2020}.
Therefore, there is no guarantee that the on-demand transport service industry
can maximize its profit by centralized planning alone,
such as via scheduling, routing, pricing, and ridesharing
\cite{Egan2016, Alonso-Mora2017, Vazifeh2018, Ramezani2018}.
To achieve economic efficiency,
we need to consider drivers' strategic behavior
when we design new models of on-demand service.

\subsection{Implementation of search strategy}
\label{sub:implementation}

Here we point out how a driver would implement a search strategy.
Picture a driver $i$ who is familiar with city traffic
and hailer and driver distributions throughout a day.
To earn more money, the driver has a plan on how much time to spend
searching different places for hailers;
the plan may vary for different time of day.
At the beginning of $i$'s shift, the driver heads to the region
where the plan allocates the most search time.
After delivering the first pickup,
the driver is likely to be in a region with less planned search time.
To avoid over-searching the current region, $i$ drives back to the preferred region.
If $i$ goes through the preferred region without a pickup,
the driver would circle around and continue the search,
as long as the total search time within the region is not too long compared with the plan.
The driver does not always search or immediately go back to the region
with the highest planned search time,
but would balance the allocation of realized search time to approximate the plan.
But when $i$ drops off at a location with very little planned search time,
the driver would directly head to a place nearby where the plan gives more search time,
since a single pass would typically suffice for the drop-off location.
Because the total search time is limited for any given shift,
the driver would not be able to perfectly implement the strategy in one shift.
But aggregated over time, the distribution of realized search time
could reasonably approximate an intended strategy.

\subsection{Markovian search strategy}
\label{sub:markov}

\begin{figure}[tb]
  \centering
  \includegraphics[width=\linewidth]{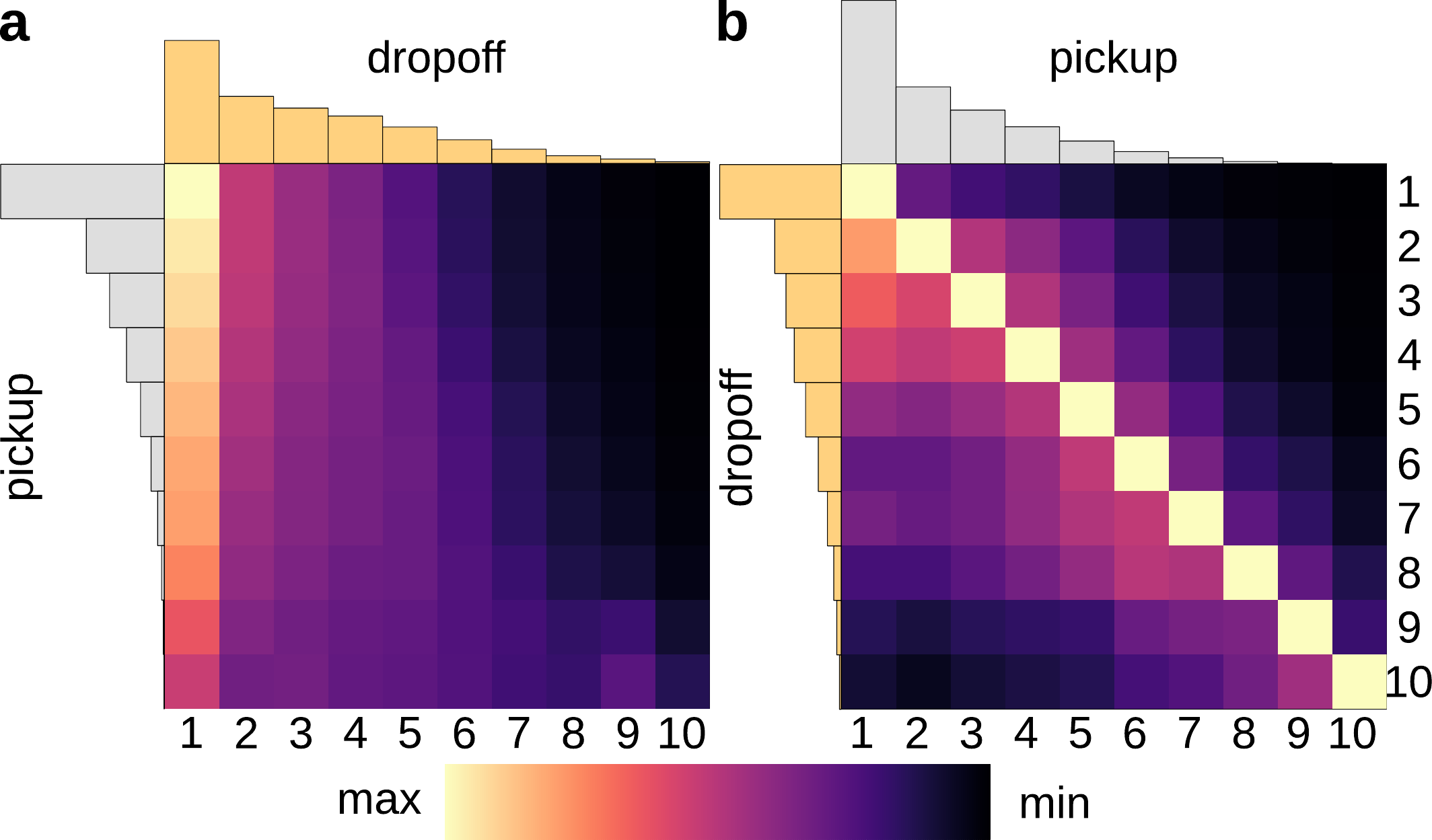}
  \caption{Markov strategy.
    Using trip records in Spring 2011 and Spring 2012, Mon-Fri 6pm--7pm.
    \textbf{a}, trip origin--destination matrix
    among 10 equal-sized groups of street segments in decreasing order of pickups.
    Rows normalized by 1-norm to show transition probability.
    Margins show pickup and drop-off counts in each group.
    \textbf{b}, search start--end matrix among the groups.
    Rows normalized by max-norm.
  }
  \label{fig:markov}
\end{figure}

An alternative formalization of driver search strategy is by Markov chain:
given the current location, the driver chooses probabilistically a neighboring location to search.
This Markovian formalization has been adopted
in game models \cite{YangH2010, HeF2015, QianXW2017},
decision models \cite{Wong2015, TangJJ2016, ZhengZ2018},
and hinted at in statistical studies \cite{LiB2011, ZhangDQ2015, Haggag2017}.
Although a Markovian search strategy may be simple to describe and implement,
it is difficult to estimate (since it involves a square matrix instead of a vector),
and does not allow a simple thermodynamic interpretation.
In this section, we give evidence of a meaningful Markovian search strategy,
and show that it is equivalent to our definition of search strategy.

If drivers are non-strategic, a null hypothesis for the Markov strategy would be random walk.
However, \cref{fig:overview}b suggests that drivers tend to move back to the area with more pickups.
We test out this hypothesis in \cref{fig:markov}.
We see that, regardless of trip origin,
locations with more pickups tend to be popular destinations as well.
Overall, the drop-off distribution is more spread out than the pickup distribution.
On the other hand, the search matrix is diagonal dominated and skews towards more popular locations.
Because search time is typically short in the PM peak \cite{ZhangRD2018phd},
most drivers do not need to drive far to find passengers,
which explains the diagonal dominance.
The skews reveal driver strategy:
drivers move back to popular pickup locations.
In particular, group 1 accounts for 42\% of pickups and 32\% of drop-offs,
and comparatively very few drivers find their next pickup in other groups;
as the drop-off location gets less popular,
the skew away from less popular groups and towards more popular ones become more prominent.
The results are similar for other time of day.

Here we show the equivalence of Markov strategy and search time allocation vector.
A Markov strategy can be represented by a search transition matrix $Q_{xy}$,
a right stochastic matrix that gives the transition probabilities
from every segment $x$ to every neighboring segment $y$ while the driver is searching for passengers.
Let $P_{xy}$ be the empirical pickup transition matrix,
i.e. row-normalized trip origin-destination matrix.
Let $p_x$ be the probability of pickup per search on a segment $x$, and $P_x = \text{diag}\{p_x\}$.
We note that $P_{xy}$ can be easily computed,
see e.g. \cref{fig:markov}a,
while $p_x$ and $Q_{xy}$ can be computed if high resolution trajectory data is available.
The equilibrium search time allocation vector $\mathbf{s}^*$ and the Markov strategy $Q_{xy}$
satisfy the equation:
\begin{equation}
  \label{eq:equivalence}
  \mathbf{s}^* = \mathbf{s}^* [P_x P_{xy} + (I - P_x) Q_{xy}]
\end{equation}
To see this, think of an ensemble of vacant vehicles, each searching for pickup.
After one time step, most continue to search on neighboring streets;
some become occupied and exit the ensemble,
and join the ensemble again after drop-off.
Note that the drivers are non-interacting.
Since we are only counting the allocation of search time,
in long-time limit each driver would have the same allocation of search time,
which equals the ensemble average at any time, and satisfies the equation as written.
Because $p_x$ and $P_{xy}$ are determined by the environment condition,
$Q_{xy}$ uniquely determines $\mathbf{s}^*$.

\subsection{Modeling demand and supply}

This paper focuses on modeling driver search strategy,
where implications of the equilibrium are validated using short time windows
during which passenger demand and driver supply can be seen as exogenous.
With this modeling choice, the only strategic component is the drivers' allocation of search time.
Another interesting problem is to study the demand and supply in on-demand mobility services.
Such models must treat passenger decisions and driver supply decisions as endogenous,
and are therefore distinct from our current study.

In \cite{ZhangRD2020dsp}, we propose a matching dynamic of passengers and taxi drivers,
and use the observed pickup data and equilibrium supply
to back out the unobserved demand at street segment level.
For on-demand services provided by transportation network companies (TNCs),
the demand and supply dynamics are necessarily different, which merit separate studies.
Perhaps surprisingly, in \cite{ZhangRD2020dsp} we show that
taxis out-perform TNCs in high-demand locations.
Future demand-supply studies should look deeper into
the differences and the interaction between on-demand services.
This would help guide regulations such as congestion charges,
to combat congestion and pollution.

\section{Materials and Methods}
\label{sec:methods}

\subsection{Taxi trip records}
\label{sub:data}

The New York City (NYC) Taxi and Limousine Commission (TLC)
started its Taxicab Passenger Enhancement Program (TPEP) in late 2008,
which collects electronic trip record of its Medallion taxis (aka yellow cabs).
TLC releases TPEP records to the public per the Freedom of Information Law of New York State.
We have gathered the records from 2009 to 2013,
the first five calendar years since TPEP devices were installed in all 13,237 Medallion taxis.
The data set contains over 870 million trips and 50,297 frequent drivers. %
Each trip record contains medallion ID (for vehicles), hack license (for drivers),
latitude, longitude and time stamp of pickup and drop-off,
trip distance, fare amounts, and other attributes.
We use the ID fields to link a taxi between consecutive trips,
and derive new attributes for use in our studies \cite{ZhangRD2020dsp,ZhangRD2018phd}.
The original and processed data are available for reuse at \cite{ZhangRD2019data}.
For many of the empirical analyses in this paper, we subset the data to short time windows
to meet the assumption of exogenous environment condition.
See \cite{ZhangRD2020dsp} for the time sampling procedure.
See \cite{ZhangRD2018phd} for more details on the data set.%

\subsection{Road network and map matching}
\label{sub:map matching}

We use OpenStreetMap (OSM) data for the public non-freeway vehicular road network in NYC.
Specifically, we include OSM ways whose \textsf{highway} tag take one of the following values:
\textsf{trunk}, \textsf{primary}, \textsf{secondary}, \textsf{tertiary},
\textsf{unclassified}, \textsf{residential}.
To make the road network strongly connected, we removed tunnels, bridges, and link roads.
The filtered OSM map has 8,928 locations and 11,458 edges.
We use Open Source Routing Machine (OSRM) to create a compressed graph of 6,001 edges.
We exploit another module in OSRM to match GPS locations to the nearest segment,
where longitudes and latitudes are transformed in Mercator projection
for isotropic local scales of distance.
The modified code is available at \url{https://github.com/rudazhan/osrm-backend}.

\subsection{Driver decision-making}
\label{sub:decision}

The behavioral rule of taxi drivers can be simply expressed as follows:
taxi drivers maximize their income by choosing their driving strategy.
We ignore drivers' exit decisions, 
and assume that individuals who drive a taxi can earn at least as much income as their cost,
i.e. their alternative income.
When this condition does not hold, rational individuals would not be driving a taxi.
Here we show that this income maximization is strategically equivalent to revenue maximization,
and we formalize each driver's decision as an optimization problem.
For the industrial organization of NYC taxis,
\cite{ZhangRD2018phd} gives a comprehensive analysis with references to rules and regulations.

The income structure of a taxi driver depends on the property rights of the taxi in use.
Owner-drivers are Medallion owners who also drive their taxis, so they have no lease to pay.
Drivers of driver-owned vehicles lease a Medallion from fleets, agents, or Medallion owners,
and either own or finance the purchase of the vehicle, at different lease costs.
Other drivers lease both the Medallion and the vehicle;
the lease may optionally include a fixed amount for gasoline surcharge since 2012-09-30.
In any type of such leases,
the driver pays a fixed amount of money either per shift which lasts 12 hours,
or per week in longer-term leases.

Taxi drivers also pay for fuel usage (if not covered by lease),
which depends on vehicle model, vehicle speed, and acceleration.
As of vehicle model,
most are gasoline or hybrid-electric vehicles:
at any time during 2009-2013, at most 23 of the 13237 Medallion taxis
use diesel or compressed natural gas vehicles.
For gasoline and hybrid light passenger vehicles operating at urban traffic speed
(16-40 km/h or 10-25 mph),
fuel consumption per hour is almost constant, see \cite{Davis2017}.
This means that fuel cost per service time can be seen as a constant for each taxicab regardless of
speed---we do not consider taxis parked by the curb with engine off actively in service.
Even without this observation,
fuel cost per service time would still be approximately constant for a driver, %
as long as the driver has consistent driving speeds and acceleration patterns.

A taxi driver earns the remaining fare and tips after paying for lease, fuel, or both.
Although vehicle maintenance is another cost to drivers who own the vehicle,
it is not relevant to the driver strategy of our interest.
Formally, the hourly income $u_i$ of driver $i$ derives from hourly trip revenue $\pi_i$,
minus hourly fuel cost $f_i$, minus amortized hourly lease payment $r_i$:
\begin{equation}
  \label{eq:income}
  u_i = \pi_i - f_i - r_i
\end{equation}  
The amortized hourly lease payment by the driver is $r_i = R_i / T_i$,
where $T_i$ denotes driver total service time during the lease term,
and $R_i$ denotes lease payment, i.e. rent of the Medallion taxicab.
Depending on the lease, $f_i$ or $r_i$ may be zero.
Since $f_i$ and $r_i$ are constant for driver $i$ in any given shift,
they do not affect the driver's driving strategy.
Thus, the objective of a driver is strategically equivalent to trip revenue $\pi_i$.%

What driving strategy can drivers use to maximize revenue?
Taxis in service are either vacant or occupied:
when vacant, drivers search the streets for hailers;
when occupied, drivers take the passengers to their destination.
Drivers can freely choose how they spend their search time over the street network.
Once they find hailers, drivers will stop the search and pick them up.
(In real life, not all taxi drivers pick up every hailer they meet.
For profitability, security, or end-of-shift concerns,
they may discriminate against hailers based on the destination, race, or other factors.
See NYC 311 records for complaints about taxis service denial.)
Taxi fare rate is set by the city government,
which may be metered or has a flat rate, depending on the destination.
Under flat rate, drivers are best off taking the fastest path.
Metered rates charge by distance or duration, based on a speed threshold,
which are typically set such that drivers have no incentive to drive slow.
Although drivers do have an incentive to take routes longer than the fastest path,
passengers typically are motivated to supervise trip duration.
In case of driver fraud, detouring is not a common strategy~\cite{Balafoutas2015}.
Thus, we assume that a taxi driver's delivery strategy
is to take passengers to their destination via the fastest path,
so trip duration between two specific locations only depends on traffic and driving speed.
We see that, consistent with the discussion in \Cref{sec:introduction},
the only strategic elements for taxi drivers are
driving speed and how they allocate their search time.%

Now we formalize drivers' search strategy.
Let $N$ be the set of taxi drivers currently in service.
Let $G = (V, E)$ be the road network within the urban area being studied,
where $V$ is the set of intersections and dead ends,
and $E$ is the set of street segments.
Street segment $x \in E$ has length $l_x$,
with traffic speed $v_x$ and taxi search speed $\tilde{v}_x$.
Define demand rate $\mu_{dxy}$ as the frequency of hailers
start hailing on segment $x$ who are going to segment $y$;
such a group of hailers have impatience $\mu_{txy} = 1 / \mathbb{E} T_{xy}$,
the reciprocal of hailer mean patience.
Within a short time interval,
environment condition $\mathcal{E} = (\mathbf{v}, \boldsymbol{\mu}_d, \boldsymbol{\mu}_t)$
can be considered as constant,
where $\mathbf{v}$ is the vector of traffic speeds,
and $\boldsymbol{\mu}_d$ and $\boldsymbol{\mu}_t$ are matrices of hailer demand and impatience.
Search strategy for driver $i$ can be defined as the spatial distribution of supply rates
$\boldsymbol{\mu}_{si}$, where $\mu_{six} = (\boldsymbol{\mu}_{si})_x$ is the frequency
at which driver $i$ enters segment $x$ as a vacant taxi.
Equivalently, driver search strategy can be defined as
the distribution of driver's search time per unit time:
\begin{equation}
  \label{eq:search-time}
  \frac{t_{six}}{t} = \frac{l_x}{\tilde{v}_x} \mu_{six}
\end{equation}
This shows that on each segment, driver search time is linearly related to driver supply rate.
Define pickup rate $\mu_{pixy}$ as the frequency at which
driver $i$ picks up passengers on $x$ going to $y$.
These attributes naturally aggregates on each segment:
$\mu_{px} = \sum_i \sum_y \mu_{pixy}$, $\mu_{sx} = \sum_i \mu_{six}$, $\mu_{dx} = \sum_y \mu_{dxy}$,
and $\mu_{tx} = 1 / \mathbb{E} T_x$.
Pickup rate can thus be expressed as a function of supply rate,
demand rate and hailer impatience: $\mu_{px}(\mu_{sx}, \mu_{dx}, \mu_{tx})$.
\cite{ZhangRD2020dsp} proposed a class of pickup models and proved that
the pickup rate functions are increasing, strictly concave, and arbitrarily differentiable,
with respect to supply rate;
for three representative models, analytical forms of the pickup rate functions are also provided.

We now relate driver search strategy with driver revenue.
Let $\Pi_{xy}$ be the revenue of a single trip from $x$ to $y$,
which only depends on traffic speeds $\mathbf{v}$.
We can write hourly revenue originated on $x$ as $\pi_x = \sum_y \Pi_{xy} \mu_{pxy}$
and average revenue of a trip originated on $x$ as $\overline{\Pi}_x = \pi_x / \mu_{px}$.
Assume patience and destination are approximately uncorrelated for hailers with the same origin,
which means $\forall x, y \in E, \mu_{tx} \approx \mu_{txy}$.
Then hailers on the same segment have an equal chance of being picked up
regardless of their destination:
\[\forall x \in E, \mu_{pxy} \propto \mu_{dxy}, \forall y \in E\]
Thus, the average revenue for a trip originated on $x$
only depends on traffic speeds and demand rates:
$\overline{\Pi}_x(\mathbf{v}, \boldsymbol{\mu}_{dx}) = \sum_y \Pi_{xy} \mu_{dxy} / \mu_{dx}$.
Since drivers are assumed not to discriminate hailers:
\[\forall i \in N, \forall x \in E, \mu_{pixy} \propto \mu_{pxy}, \forall y \in E\]
Driver revenue originated on a segment $\pi_{ix} = \sum_y \Pi_{xy} \mu_{pixy}$ can thus be written as
$\pi_{ix} = \sum_y \Pi_{xy} \mu_{pxy} \mu_{pix} / \mu_{px} = \overline{\Pi}_x \mu_{pix}$.
Since each pass of a vacant taxi has an equal chance of picking up a hailer regardless of the driver:
\begin{equation} \label{eq:prop-pickup-search}
  \forall x \in E, \mu_{pix} \propto \mu_{six}, \forall i \in N
\end{equation}
We have $\pi_{ix} = \overline{\Pi}_x \mu_{pix} = \overline{\Pi}_x \mu_{px} \mu_{six} / \mu_{sx}$.
Driver hourly trip revenue can thus be expressed with explicit function dependency as:
\begin{equation*}
  \pi_i = \sum_{x \in E} \pi_{ix} = \sum_{x \in E} \overline{\Pi}_x (\mathbf{v},
  \boldsymbol{\mu}_{dx}) \mu_{px}(\mu_{sx}, \mu_{dx}, \mu_{tx}) \frac{\mu_{six}}{\mu_{sx}}
\end{equation*}

A more analytically convenient definition of driver search strategy
is driver's allocation of service time.
Service time $t_{ix} = t_{six} + t_{pix}$ is the total time driver $i$ spends
searching and delivering trips originated on $x$ during a period of time $t$.
The rationale of using service time distribution as driver search strategy
instead of supply rate or search time is that:
service time is a conserved quantity and identical for all drivers;
meanwhile, service time is monotonic in supply rate
and preserves properties of the pickup rate function.
Let $t_{xy}$ be the trip duration from $x$ to $y$, which only depends on traffic speeds $\mathbf{v}$.
The average duration of a trip originated on $x$ is
$\overline{t}_x(\mathbf{v}, \boldsymbol{\mu}_{dx}) = \sum_y t_{xy} \mu_{dxy} / \mu_{dx} =
\sum_y t_{xy} \mu_{pxy} / \mu_{px}$,
with reasoning similar to average trip revenue $\overline{\Pi}_x$.
The proportion of time driver $i$ spends delivering trips originated on $x$ is thus
$t_{pix} / t = \sum_y t_{xy} \mu_{pixy} = \sum_y t_{xy} \mu_{pxy} \mu_{pix} / \mu_{px} =
\overline{t}_x \mu_{pix} = \overline{t}_x \mu_{px} \mu_{six} / \mu_{sx}$,
with reasoning similar to $\pi_{ix}$.
Together with \cref{eq:search-time},
the proportion of service time driver $i$ allocates on $x$ can thus be written as:
\begin{equation}
  \label{eq:service-time}
  s_{ix} = \frac{t_{six} + t_{pix}}{t} = \left( \frac{l_x}{\tilde{v}_x} +
    \overline{t}_x \frac{\mu_{px}}{\mu_{sx}} \right) \mu_{six}
\end{equation}
This shows that on each segment, driver service time is also linearly related to driver supply rate:
$\forall x \in E, s_{ix} \propto \mu_{six}, \forall i \in N$.
From \cref{eq:service-time}, service time on a segment
$s_x = \mu_{sx} l_x / \tilde{v}_x + \mu_{px} \overline{t}_x$.
With pickup rate function $\mu_{px}(\mu_{sx}, \mu_{dx}, \mu_{tx})$
and constant environment condition $\mathcal{E}$,
pickup rate is implicitly a function of service time: $\mu_{px}(s_x, \mathcal{E})$.
Each taxi driver must allocate all the service time among the street segments:
$\sum_x t_{ix} = t$, or equivalently $\sum_x s_{ix} = 1$.
The search strategy of taxi driver $i$ is thus $\mathbf{s}_i \in S_i$,
where the strategy space $S_i = \Delta^{|E|-1}$,
a simplex of dimension one less than the number of segments.
Now we can formalize the optimization problem of a taxi driver:
\begin{equation}
  \label{eq:game}
  \begin{aligned}
    & \text{maximize} && \sum_{x \in E} \overline{\Pi}_x(\mathbf{v},
    \boldsymbol{\mu}_{dx}) \mu_{px}(s_x, \mathcal{E}) \frac{s_{ix}}{s_x} \\ %
    & \text{subject to}  && \mathbf{s}_i \ge 0 \\
    &                    && \mathbf{s}_i \cdot \mathbf{1} = 1
  \end{aligned}
\end{equation}

Now we prove that pickup rate $\mu_{px}(s_x, \mathcal{E})$ is also increasing,
strictly concave, and arbitrarily differentiable with respect to $s_x$,
so that \cref{eq:game} satisfies the requirements of multi-market oligopoly \cite{ZhangRD2020game}.
With constant environment condition $\mathcal{E}$,
the implicit function can be abstracted to $z = a x + b y$,
where $z = s_x$, $x = \mu_{sx}$, $y = \mu_{px}$, $a = l_x / \tilde{v}_x$, and $b = \overline{t}_x$;
$y(x)$ is increasing, strictly concave, and arbitrarily differentiable,
while $a, b > 0$ are constants.
Our proposition is thus equivalent to: $y(z)$ is also increasing, strictly concave,
and arbitrarily differentiable.
Differentiability is simply preserved by the linear relation.
Since $z(x) = a x + b y(x)$ is increasing, its inverse $x(z)$ is thus also increasing;
by composition, $y(z) = y(x(z))$ is also increasing.
By implicit differentiation, $\mathrm{d}y/\mathrm{d}z = y'(x) / (a + b y'(x))$,
and thus $\mathrm{d^2}y/\mathrm{d}z^2 = a y''(x) / (a + b y'(x))^3$.
Since $y'(x) > 0$ and $y''(x) < 0$, $y''(z) < 0$, which means $y(z)$ is also strictly concave.

\section*{Acknowledgment}
\label{sec:Acknowledgment}

The authors thank Henry S. Farber of Princeton University and Abhishek Nagaraj of UC Berkeley
for sharing NYC taxi trip records, and thank Ketan Savla, Sami F. Masri, Juan Carrillo,
Matthew Kahn, Hashem Pesaran, and Geert Ridder of USC for helpful comments.

\bibliographystyle{myIEEEtran} %
\bibliography{new.bib}

\begin{IEEEbiography}[{\includegraphics[width=1in,height=1.25in,clip,keepaspectratio]{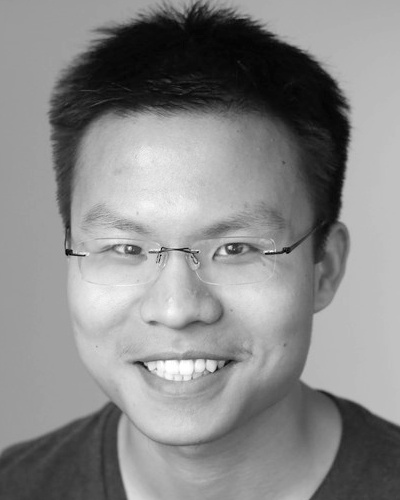}}]
  {Ruda Zhang} received his B.E. degree from Peking University, Beijing, China, in 2012,
  and M.A. degree in economics and Ph.D. degree in civil engineering
  from University of Southern California, Los Angeles, CA, in 2018.
  He is currently a postdoctoral fellow at the Statistical and Applied Mathematical Sciences Institute.
  His research interests include urban systems, transportation, sensor data analysis, and game theory.
\end{IEEEbiography}

\begin{IEEEbiography}[{\includegraphics[width=1in,height=1.25in,clip,keepaspectratio]{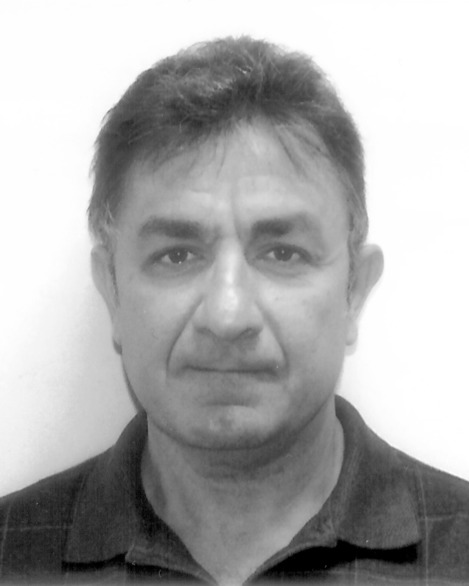}}]
  {Roger Ghanem} is the Gordon S. Marshall Professor of Engineering Technology
  in the Department of Civil \& Environmental Engineering at the University of Southern California.
  He received his Bachelor of Engineering degree from the American University in Beirut in 1984
  and his Masters and PhD degrees from Rice University in 1985 and 1989, respectively.
  Professor Ghanem's research is in the area of computational stochastic mechanics and
  uncertainty quantification with focus on coupled, heterogeneous and multiscale systems.
  He is fellow of USACM, WCCM, EMI, AAAS, and SIAM.
\end{IEEEbiography}

\end{document}